\documentclass[aps,prl,twocolumn,showpacs]{revtex4-1}

\usepackage{amsmath,amssymb,graphicx,graphics,bm}
\usepackage{hyperref}
\usepackage{graphics,graphicx}
\usepackage{amsmath,amssymb}

\begin{document}

\title{Another normality is possible. \\  Distributive transformations and emergent Gaussianity}

\author{Massimiliano Giona, Chiara Pezzotti and Giuseppe Procopio}
\email[corresponding author:]{massimiliano.giona@uniroma1.it}
\affiliation{Dipartimento di Ingegneria Chimica, Materiali, Ambiente La Sapienza Universit\`a di Roma\\ Via Eudossiana 18, 00184 Roma, Italy}

\date{\today}

\begin{abstract}
A distributional route to Gaussianity, associated with the concept of
Conservative Mixing Transformations in ensembles of random vector-valued variables, is proposed.
This route is completely different from  the additive mechanism characterizing the application of Central Limit Theorem, as it is
based on the iteration of a random transformation preserving  the ensemble variance.
Gaussianity emerges as a ``supergeneric'' property of ensemble statistics, in the case the energy constraint 
is quadratic in the norm of the variables. This result puts in a  different  light the occurrence
of equilibrium Gaussian distributions in kinetic variables (velocity, momentum),
as it shows mathematically that, in the absence of any other dynamic mechanisms, almost Gaussian
distributions stems from the low-velocity approximations of the physical conservation principles.
Whenever, the energy constraint is not expressed in terms of quadratic functions (as in the relativistic case), 
the J\"uttner distribution is recovered from CMT.
\end{abstract}

\maketitle

The concept of normal distribution is so central in probability theory and physics that, as observed by Mark Kac, partially in jest (quoting Henry Poincar\'e), it is not clear
in the scientific community whether it comes from a mathematical
property or it is a law of nature \cite{kaclibro}.
As well known, the main route to normal distributions stems from the 
Central Limit Theorem (CLT) \cite{cltgen1,cltgen2}. There is an entire galaxy of
different versions of CLT,  either relaxing the assumptions  and the convergence properties,
 or modifying the way  the summation is  performed \cite{cltgen3,cltgen4}.
Similarly, there are parallel extensions of CLT, initiated by the
work of C. Stein,   applied and generalized in several directions, e.g. for sums of dependent random
variables \cite{stein2}.

Besides these mathematical subtleties and chiselings, of 
utmost value in probability theory and statistics, the core
of CLT -  finding a widespread application in  statistical physics 
for interpreting the phenomenological
occurrence of  normal distributions \cite{statphys1} -
lies in its simplest version: given a sequence of independent and
identically distributed (iid) random variables $\{ X_n \}_{n=1}^\infty$,
possessing zero mean and bounded variance $\sigma$, the random
variable $Y_N= \sum_{n=1}^N X_n/\sigma_{Y,N}$,  corresponding
to their  sum normalized  to unit variance, 
(as, by independence, $\sigma_{Y_N}^2=N \, \sigma^2$), converges  weakly for $N \rightarrow
\infty$   to the normal  distribution, so that its limit probability
density function (pdf) is given  by $p_Y(y)= e^{-y^2/2}/\sqrt{2 \, \pi}$,
(henceforth the normal distribution is  indicated with  the symbol ${\mathcal N}(0,1)$.).
It is important to observe that the weak convergence, i.e., the convergence  in distribution,
does not prevent the potential occurrence of anomalies as regards the higher order moments  $m_{Y_N}^{(n)}=\langle Y_N^n \rangle$,
for $n > 2$ \cite{gionaklages}.

CLT  explains in a simple  and elegant way the
phenomenon of molecular diffusion addressed by A. Einstein in 
relation to the phenomenon of Brownian motion \cite{einstein}: the
molecular   motion, or the kinematics of micrometric particles driven by thermal
fluctuations,  can be interpreted as the resultant of the
superposition  of independent (uncorrelated) displacements that, provided
the boundedness of their variances, unavoidably leads by mathematical principles (i.e,. by enforcing CLT) 
to the occurrence in the long-time limit of the parabolic diffusion equation.
Thus  invoking CLT,  the diffusion equation
follows, just by randomness, independence and  the smallness of the perturbations.
By the internal coherence of CLT, this interpretation provides  also
the criteria  for  assessing the failure of 
the parabolic approximation:
(i) lack of independence, leading to subdiffusion, as in the case of  the  
random
motion in disordered systems and fractal media \cite{fractal1,fractal2},
or  (ii) unbounded lower-order moments, associated with superdiffusive phenomena
 such as L\'evy walks \cite{super1,super2}, that are 
strongly related to the extended
version of CLT (usually referred to as  the generalized CLT) and to the
 L\'evy theory of $\alpha$-stable distributions
\cite{levy,alpha}. 

In its basic ``machinery'',  CLT can be  viewed as a mathematical superpositional
route to ${\mathcal N}(0,1)$, in
which normality is achieved by addition  of independent
contributions, properly rescaled.
In this Letter we want to present another, completely different, mechanism
 leading to normality, in which the Gaussian distribution
follows as a consequence of a distributional dynamics involving the
binary interaction of iid random variables subjected to some
conservation principles and  possessing an arbitary initial
distribution.
In the present case, the conservation principles 
are  the fundamental conservation principles of physics,
that apply   both in the classical and in the quantum world,  namely the
conservation of momentum and energy, both for matter-matter and
matter-radiation interactions. This  finds in the  relativistic space-time
formalism
its maximum compactness and elegance,
as it implies, for any physical interactions,  
the overall 4-momentum $p^{\mu}$, $\mu=1,..,4$,
$p^{\mu}= \left ( {\bf p}, \frac{E}{c} \right )$
is  conserved, where ${\bf p}=(p^1,p^2,p^3)$
is the momentum,  $E$ the energy and $c$ the speed of light {\em in vacuo} \cite{relativity}

Specifically, we introduce and analyze  the distributional route to
normality associated with the concept of {\em Conservative
Mixing Transformation} (CMT) of random ensembles, for which
the occurrence of limit Gaussian distributions is entirely
entitled to the  energy representation, as discussed in the
analysis of low-velocity vs relativistic constraints.

CMT's originate from the stochastic generalization  of the Boltzmannian 
description of binary collisions in diluted gases
\cite{boltz1,boltz2}. Their introduction
  does not   simply represent  a stochastic transposition of the Boltzmannian 
kinetic theory.
From the statistical analysis of CMT it is possible to derive the extreme genericity (as  explained 
in  the remainder) of the occurrence of ``almost'' Gaussian velocity distributions at equilibrium, and their connection with
the nature and representation of fundamental conservation principles.
Moreover, they provide a formal stochastic setting amenable to be extended 
to any physical impulsive interaction mechanism,
and involving arbitrary conservation principles.
In the present Letter, CMT is analyzed beyond the classical low-velocity conditions in the relativistic case. Although the Boltzmannian kinetic theory
leads to Gaussian  distributions at equilibrium, no one, to the best
of our knowledge, has extended  and generalized the mechanism of binary
interactions in the form of a universal stochastic route to Gaussianity,
which is the scope of CMT's and of the present Letter.

The structure of this Letter is organized as follows. We formalize the
concept of CMT, the nature of the physical constraints, and the generic emergence of almost Gaussian distributions.
Subsequently,  we address how Gaussianity can be broken, by the assumption of an energy conservation law different from a quadratic one.

Consider an ensemble of $N$ random  vector-valued variables ${\mathcal E}=\{ {\bf z}_h \}_{h=1}^N$, with ${\bf z}_h=(z_{h,1},\dots,z_{h,d}) \in {\mathbb R}^d$, $d=1,2,\dots$,
and let $\Sigma_{d,N}$ the space of all the $N$-dimensional
ensembles of $d$-dimensional random  variables over the field  ${\mathbb R}$.
Each ${\bf z}_h$ can be referred to as the state vector of an element of the ensemble.
A Mixing Transformation ${\mathcal M} : \Sigma_{d,N} \rightarrow \Sigma_{d,N}$ (acronym MT)
is a transformation of  the ensemble ${\mathcal E}$, into an ensemble ${\mathcal E}^\prime={\mathcal M}({\mathcal E})=
\{ {\bf z}_h^\prime \}_{h=1}^N$, defined  by a binary operation
amongst randomly selected elements of the ensemble. It can be defined in  the following way:
\begin{enumerate}
\item let $\boldsymbol{\phi}({\bf z}_1,{\bf z}_2;{\bf r})$,
$\boldsymbol{\psi}({\bf z}_1,{\bf z}_2;{\bf r}) : {\mathbb R}^d \times {\mathbb R}^d \times \partial S_d \rightarrow {\mathbb R}^d$ be two random functions, 
depending on a 
 random vector ${\bf r}$, $|{\bf r}|=1$ defined on the surface of the  $d$-dimensional  unit sphere $\partial S_d$,
by the pdf $g({\bf r})$;
\item select randomly two elements $\alpha, \beta \in (1,\dots,N)$ with $\beta \neq \alpha$;
\item the transformed ensemble $\{ {\bf z}_h^\prime \}_{h=1}^N$ is given by
\begin{eqnarray}
{\bf z}_\alpha^\prime  & =  & \boldsymbol{\phi}({\bf z}_\alpha,{\bf z}_\beta;{\bf r})
\nonumber \\
{\bf z}_\beta^\prime  & =  & \boldsymbol{\psi}({\bf z}_\alpha,{\bf z}_\beta;{\bf r})
\label{eq2} \\
{\bf z}_h^\prime & = & {\bf z}_h \,, \qquad \mbox{for} \;\; h \neq \alpha,\beta
\nonumber
\end{eqnarray}
\end{enumerate}
Whenever it is not conceptually necessary, the  explicit dependence of ${\boldsymbol \phi}$ and ${\boldsymbol \psi}$ on ${\bf r}$ will be omitted.

The concept of MT's so defined is too general for physical
applications, and constraints 
on the random functions $\boldsymbol{\phi}$, $\boldsymbol{\psi}$,
should  be introduced.

A Conservative Mixing Transformation  is a MT, for which $N_c$ functions $f_h({\bf z}):  {\mathbb R}^d \rightarrow {\mathbb R}$
are defined, such that the transformations $\boldsymbol{\phi}({\bf z}_1,{\bf z}_2)$  and $\boldsymbol{\psi}({\bf z}_1,{\bf z}_2)$ 
satisfy the constraints
\begin{equation}
f_h({\bf z}_1^\prime)+ f_h({\bf z}_2^\prime)=f_h({\bf z}_1)+f_h({\bf z}_2) \, , \qquad h=1,..,N_c
\label{eq3}
\end{equation}
where ${\bf z}_1^\prime= \boldsymbol{\phi}({\bf z}_1,{\bf z}_2)$, ${\bf z}_2^\prime= \boldsymbol{\psi}({\bf z}_1,{\bf z}_2)$,
for any ${\bf z}_1, {\bf z}_2 \in {\mathbb R}^d$, and ${\bf r} \in \partial S_d$.

The CMT's of  physical relevance are those satisfying the $N_c=d+1$ conservation laws
\begin{eqnarray}
f_{h}({\bf z}) & = & z_h  \, , \quad h=1,\dots,d \label{eq4} \\
f_{d+1}({\bf z}) & = & e(|{\bf z}|) \label{eq5}
\end{eqnarray}
where $e(|{\bf z}|)$ is a non negative function solely of the norm $|{\bf z}|$ of  ${\bf z}$,
representing, modulo a multiplicative factor, the energy function (kinetic energy) of the element.
Eqs. (\ref{eq4}) and (\ref{eq5}) correspond mathematically to the consevation of momentum
and energy. 
Several observation, follows  from this setting. (i) The concept of
random ensembles involves a finite number $N$ of elements, and should not be confused with
Gibbsian ensembles or with other collective groupings (such as in the replica method) \cite{gibbs,replica}.
A ``random ensemble''  in the CMT-theory means simply a system of random variables, in the sense that
(i) their initial conditions are randomly chosen for each element, and that their evolution is subjected to
random laws. 
Correspondingly, the ensemble average of any function $q({\bf z})$ of the state vector ${\bf z}$, is simply
expressed by $\langle q({\bf z}) \rangle = \frac{1}{N} \sum_{h=1}^N g({\bf z}_h)$, and consequently a single
pdf $p_z({\bf z})$ is associated with the ensemble.
(ii) The definition of MT proposed above is ``event-based'', in the meaning that each transformation
${\mathcal M}$ involves solely a single binary event  modifing the statistical properties of
the ensemble.
If ${\mathcal E}^0=\{ {\bf z}_h^{(0)}\}_{h=1}^N$ is the initial ensemble,
assume  $\langle {\bf z}^{(0)} \rangle =0$,
and $\langle e(|{\bf z}^{(0)})|\rangle =E_0 < \infty$.

Let $({\boldsymbol \phi}^{-1},{\boldsymbol \psi}^{-1})$ be the inverse transformation of $({\boldsymbol \phi},{\boldsymbol \psi})$,
i.e., ${\boldsymbol \phi}^{-1}({\boldsymbol \phi},{\boldsymbol \psi})=  {\boldsymbol \psi}^{-1}({\boldsymbol \phi},{\boldsymbol \psi})= \mbox{Id.}$, and set
\[     
J_{\phi,\psi}({\bf z}_1^\prime,{\bf z}_2^\prime)| = \left |
\frac{\partial \left ({\boldsymbol \phi}({\bf z}_1,{\bf z}_2),{\boldsymbol \psi}({\bf z}_1,{\bf z}_2) \right )}{\partial ({\bf z}_1,{\bf z}_2)} \right |_{\begin{array}{l}
                                                                                                {\bf z}_1={\boldsymbol \phi}^{-1}({\bf z}_1^\prime,{\bf z}_2^\prime) \\
                                                                                                {\bf z}_2 = {\boldsymbol \psi}^{-1}({\bf z}_1^\prime,{\bf z}_2^\prime)
                                                                                                 \end{array} }
\]
for the Jacobian determinant of the transformation.
The statistical evolution of a CMT can be described as follows. 
If pdf $p_z({\bf z})$ is the ensemble pdf for ${\mathcal E}$ and ${\mathcal E}^\prime={\mathcal M}({\mathcal E})$, the  pdf $p_z^\prime({\bf z}^\prime)$ of ${\mathcal E}^\prime$,
averaged over the probability measure of ${\bf r}$, is expressed by 
\begin{equation}
p_z^\prime({\bf z}^\prime)= \frac{1}{2} \, \int_{{\mathbb R}^d} \left [ \pi^\prime({\bf z}^\prime,{\bf z}_1)+ \pi^\prime({\bf z}_1,{\bf z}^\prime) \right ]
\, d {\bf z}_1
\label{eq6}
\end{equation}
where
\begin{equation}
 \pi^\prime({\bf z}_1^\prime,{\bf z}_2^\prime) = \int_{\partial S_n} \frac{p_z({\boldsymbol \phi}^{-1}({\bf z}_1^\prime,{\bf z}_2^\prime;{\bf r})) 
 p_z({\boldsymbol \psi}^{-1}({\bf z}_1^\prime,{\bf z}_2^\prime;{\bf r})) g({\bf r})d {\bf r}}{|J_{\phi,\psi}({\bf z}_1^\prime,{\bf z}_2^\prime)|} 
\label{eq7}
\end{equation}
The independence of ${\bf z}_1$ and ${\bf z}_2$, expressed by the factorization
of the two densities in the integrand at the r.h.s. of eq. (\ref{eq7})
stems from the random selection rule (point 2.) in the
definition of a MT.
Consider for the energy function the expression,
\begin{equation}
e(|{\bf z}|)= |{\bf z}|^2
\label{eq8}
\end{equation}
corresponding to the classical  form for the kinetic energy 
(for identical particles) 
In this case, the transformations  ${\boldsymbol \phi}({\bf z}_1,{\bf z}_2;{\bf r})$, and ${\boldsymbol \phi}({\bf z}_1,{\bf z}_2;{\bf r})$
can be chosen as
\begin{equation}
\left \{
\begin{array}{l}
{\boldsymbol \phi}({\bf z}_1,{\bf z}_2;{\bf r})  = (1-\lambda) \,  {\bf z}_1+ \lambda \, {\bf z}_2 + \alpha_\lambda({\bf z}_1,{\bf z}_2) \, {\bf r} \\
{\boldsymbol \psi}({\bf z}_1,{\bf z}_2;{\bf r})  = \lambda \,  {\bf z}_1+ (1-\lambda) \, {\bf z}_2 - \alpha_\lambda({\bf z}_1,{\bf z}_2) \, {\bf r}
\end{array}
\right .
\label{eq9}
\end{equation}
where $\lambda \in [0,1]$ is a parameter, and $\alpha_\lambda({\bf z}_1,{\bf z}_2)$
is defined  to fulfil  eqs. (\ref{eq3}), (\ref{eq5}). For $\lambda=0$, $\alpha_\lambda=-({\bf z}_1-{\bf z}_2) \cdot {\bf r}$ (where
``$\cdot$'' indicates the Euclidean scalar product), for $\lambda=1/2$, $\alpha_\lambda= |{\bf z}_1-{\bf z}_2|/2$.
Consider $\lambda=1$, as the other cases are equivalent. It is easy to check that
$|J_{\phi,\psi}|=1$, so that, if  $p_z^*({\bf z})$ is the equilibrium distribution, $\pi^\prime({\bf z}_1^\prime,{\bf z}_2^\prime)
=p_z^*({\bf z}_1^\prime) \, p_z^*({\bf z}_2^\prime)$, and eq. (\ref{eq7})
becomes
\begin{eqnarray}
p_z^*({\bf z}_1^\prime) \, p_z^*({\bf z}_2^\prime)  & =  & 
\int_{\partial S_d} p_z^* \left (  {\bf z}_1^\prime - ({\bf z}_1^\prime-{\bf z}_2^\prime) \cdot {\bf r} \, {\bf r} \right ) \nonumber \\
& \times & p_z^* \left (  {\bf z}_1^\prime + ({\bf z}_1^\prime-{\bf z}_2^\prime) \cdot {\bf r}
 \, {\bf r} \right ) \, g({\bf r}) \, d {\bf r}
\label{eq10}
\end{eqnarray}
Since,
\[
| {\bf z}_1^\prime- ({\bf z}_1^\prime-{\bf z}_2^\prime)
\cdot {\bf r} \, {\bf r} |^2+ | {\bf z}_1^\prime- ({\bf z}_1^\prime-{\bf z}_2^\prime) 
\cdot {\bf r} \, {\bf r} |^2 = |{\bf z}_1^\prime|^2+ |{\bf z}_2^\prime|^2
\]
the solution of eq. (\ref{eq10}), is given by the Gaussian
\begin{equation}
p^*({\bf z})= A e^{-\beta \, |{\bf z}|^2}
\label{eq11}
\end{equation}
where the parameter $\beta$ depends on the initial ensemble variance, and
$A$ is the normalization constant.
In deriving eq. (\ref{eq11}) we have not made use of the property of $g({\bf r})$, and therefore,
eq. (\ref{eq11}) is valid for any statistical structure of ${\bf r}$, providing that it gives rise to
a steady and unique equilibrium distribution. There is only an 
exception to this  property, represented
by {\em non-dispersive} random transformations, defined as follows.
The transformations
${\boldsymbol \phi}$ and ${\boldsymbol \psi}$ of a CMT sa\-tisfying
eqs. (\ref{eq3})-(\ref{eq4}) are {\em non-dispersive} if for 
all the allowable values of ${\bf r}$
${\boldsymbol \phi}_h({\bf z}_\alpha,{\bf z}_\beta)$ is either $z_{\alpha,h}$, or $z_{\beta,h}$, $h=1,\dots,d$ (and complementarily for ${\boldsymbol \psi}$).
This means that no real mixing amongst the entries  of the state variables occur but, componentwise,
the two transformations are either the identify or simply determine an exchange of values amongst the elements.
An example  of non-dispersive CMT using eq. (\ref{eq9}) with $\lambda=0$  
occurs for $d=2$, where ${\bf r}=(\cos \varphi, \sin \varphi)$,
$\varphi \in [0,2 \,\pi)$, in the case $g({\bf r}) \, d {\bf r}= g_\varphi(\varphi) \, d \varphi$ is any atomic
distribution of the form $g_\varphi(\varphi)= (1-\mu) \, \delta (\varphi)+ \mu \, \delta(\varphi-\pi/2)$, $\mu \in (0,1)$.
When $\varphi=0$, ${\bf z}_\alpha^\prime=(z_{\beta,1},z_{\alpha,2})$, ${\bf z}_\beta^\prime=(z_{\alpha,1},z_{\beta,2})$,
while for $\varphi=\pi/2$, ${\bf z}_\alpha^\prime=(z_{\alpha,1},z_{\beta,2})$, ${\bf z}_\beta^\prime=(z_{\beta,1},z_{\alpha,2})$.

We can state the following result with reference to the CMT transformations with energy constraint given by eq. (\ref{eq9}):
{\em for any initial distribution with $E_0>0$, for any $\lambda \in [0,1]$,  and for almost all the probability measures 
of the random vector ${\bf r}$,  the ensemble pdf of the CMT converges
in the limit for $N \rightarrow \infty$ to a Gaussian distribution.}

This stems for $\lambda=0$ from eq. (\ref{eq10}), and  similarly 
for any $\lambda \in [0,1]$ from the analogous property.
The limit for $N \rightarrow \infty$ is here introduced in order to consider an infinite ensemble, for
which a smooth and continuous probabilistic characterization (i.e., the existence of a smooth pdf) can be applied.
For any large but  finite $N$, the resulting limit distribution is still accurately approximated
by a Gaussian distribution, apart for the asymptotic tails, that necessarily vanishes to zero, as for any
finite $N$,  the support of the limit pdf should be compact,  simply
because of energy conservation and of the initial assumption of finite $E_0$.

It is interesting to compare the classical CLT route to Gaussianity with the  emergence of it stemming from
the iterative application of CMT's.  In the CLT route, normality is an emergent property of the procedure of summing
a large number of independent contributions. The existence of a limit density stems
from a renormalization procedure, of rescaling the summation by removing its mean and normalizing its variance.
In this sense the observation of Jona-Lasinio \cite{jona} on the strong analogy between CLT and the Renormalization Group
of quantum field theory is acute and cogent. In the CMT route,  the process is purely distributional
within a closed system (the ensemble) of vector-valued  random variables. Using eq. (\ref{eq8}),  both the mean and the variance of 
the  ensemble distribution are conserved in the process. There is no renormalization procedure in this approach.
Moreover, in finite ensembles the convergence towards the Gaussian distribution is only approximate (albeit arbitrarily accurate for large $N$).
This is an important, physical property, as it  resolves
 the unphysical tails for arbitrarily large $|{\bf z}|$ in real finite systems.

Moreover, what is remarkable in the distributional route to normality expressed by CMT is its genericity. While
in CLT the emergent Gaussian behavior holds for any distribution of iid random variables (with bounded mean and variance),
in the iteration of CMT it emerges: (i) for generic initial
ensemble distributions, (ii) for generic transformations ${\boldsymbol \phi}$, ${\boldsymbol \psi}$
at least expressed by eq. (\ref{eq8}); (iii) for an almost generic statistical nature of the random variable ${\bf r}$,
i.e. apart from the very peculiar (and physically irrelevant) non-dispersive transformations.
For this reason it is legitime to refer to this qualitative behavior as the ``supergenericity'' of the
CMT distributional route to Gaussianity, in analogy with the concept of ``superuniversality'' coined for
phase transitions (see \cite{superuniv} and references therein).
In a physical perspective, ``supergenericity'' is the mathematical counterpart of a thermodynamic
principle at work in CMT, determining a statistical emergent behavior that is completely independent either
of the details on the initial state or of the details on the transformations involved at a microscopic level.

\begin{figure}[!]
\includegraphics[width=6cm]{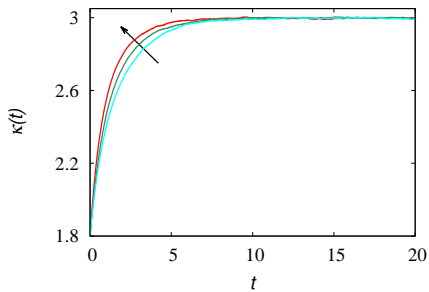}
\caption{Kurtosis   $\kappa(t)$ of the marginal ensemble distribution (for the first entry $z=z_{h,1}$ of ${\bf z}_h$) vs the normalized operational
 time $t=n/N$.
The arrow indicates first the $d=2$ case (either with $\lambda=0$ or $\lambda=1/2$, that practically coincide), then the $d=3$ case at $\lambda=0$, and 
finally
the $d=3$ case at $\lambda=1/2$.}
\label{Fig1}
\end{figure}

To  make an example, figure \ref{Fig1} depicts the evolution
of the kurtosis $\kappa(t)$ vs the  operational time $t=n/N$,
(where $n$ is the number of CMT operations, and $N$ is the ensemble size)
for the first ensemble entry $z_{h,1}$ at its convergence to
the Gaussian limit $\kappa=3$ for $d=2,\,3$ and for two different
values of $\lambda$ entering eq. (\ref{eq9}). In this case $N=10^6$, and the
entries of the initial ensamble are uniformly distributed with zero
mean and unit variance. 
All the stochastic simulations refer to a random perturbation ${\bf r}$
uniformly distributed in $\partial S_d$.
The equilibrium probability density for a larger ensemble $N=10^9$,
starting from the same initial distribution is depicted in figure \ref{Fig1}
at $d=3$, for $\lambda=0, \, 1/2$ .

\begin{figure}[!]
\includegraphics[width=6cm]{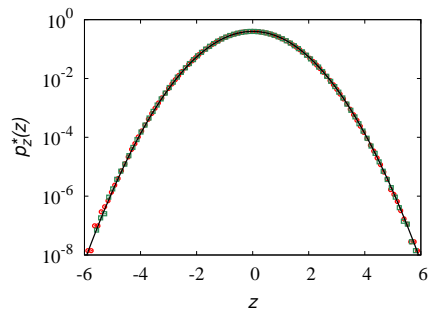}
\caption{Equilibrium probability density functions for the third entry  $z=z_{h,3}$ of ${\bf z}_h$, starting  from
the same initial conditions as in figure \ref{Fig1} for the CMT dynamics eq. (\ref{eq8}). Symbols refer to stochastic
simulations over an ensemble of $N=10^9$ elements: ($\circ$): $\lambda= 0$, ($\square$): $\lambda=1/2$.
The solid line represents the normal distribution.}
\label{Fig2}
\end{figure}

The emergence of a limit Gaussian density in CMT is {\em physics-dependent}.  
Assuming the constraints eq. (\ref{eq4}), this entirely depends on
the form of the energy constaint eq. (\ref{eq5}).
 In the presence of energy functions $e(|{\bf z}|)$ different from 
eq. (\ref{eq8})  the stationary pdf for the entries of ${\bf z}$ is different from the Gaussian.  This can be illustrated by means of a simple
example of physical relevance.
Consider the  relativistic extension \cite{dunkel1,dunkel2,giona_rel}, and consequently
the energy function $e(|{\bf z}|)$ given by
\begin{equation}
e(|{\bf z}|)= \sqrt{ |{\bf z}| \, c^2 + m^2 c^4}
\label{eq12}
\end{equation}
corresponding to the relativistic energy of particle of mass $m$, provided that
${\bf z}$ represents its momentum, while keeping the linear constraints eq. (\ref{eq4}). Set $m=1$, $c=1$ a.u.

\begin{figure}[!]
\includegraphics[width=6cm]{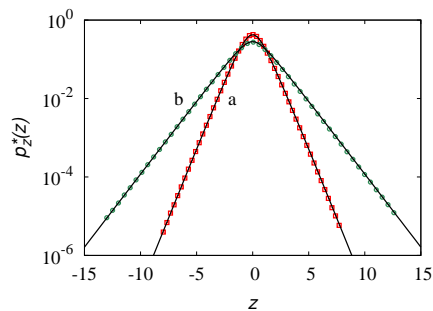}
\caption{Equililibrium probability density functions  $p_z^*(z)$ for the first entry  $z=z_{h,1}$ of ${\bf z}$ for $d=3$
in the presence of the energy function eq. (\ref{eq12}). Symbols are the results of stochastic simulations
of the CMT with $E_0=2$ (symbols $\square$), and $E_0=3$ (symbols $\circ$). Lines (a) and (b) correspond to
 the J\"uttner distribution eq. (\ref{eq13}) at two values of the parameter $\beta$.}
\label{Fig3}
\end{figure}

CMT transformations can be applied on equal footing to this case, adopting
for the transformations ${\boldsymbol \phi}$, ${\boldsymbol \psi}$ the
functional form eq. (\ref{eq9}).  The case $\lambda=0$ is considered,
where the group $\alpha_\lambda$ is defined in order to
account for the structure of the energy function  eq. (\ref{eq12}).
Figure \ref{Fig3} depicts the asymptotic (equilibrium) densities at $d=3$
for a generic entry of ${\bf z}_h$, at two different values of $E_0$. 
Data refer to an ensemble size of $N=10^7$, and the initial conditions
are of impulsive nature. Each initial ${\bf z}_{h,k}^{(0)}$ attains
with equal probability the values $\pm a$, where  $a=\sqrt{(E_0^2-1)/3}$.

Deviation for the Gaussian behaviour is sensible, and the simulation
data converge to the J\"uttner distribution 
\begin{equation}
p_z^*(z)=A e^{- \beta \,\sqrt{z^2 \, c^2 +m^2 \, c^4}}
\label{eq13}
\end{equation}
where $A$ is the normalization constant, and the parameter $\beta$
is determined by the initial value $E_0$ of the ensemble average of $e(|{\bf z}|)$.
The detailed analysis of the relativistic case is marginal in the
present discussion, and it will thoroughly developed elsewhere.
What is significant for the scope of this Letter is that Gaussianity in CMT's is a consequence of the physical assumptions on the energy constraint.

To conclude, CMT's provide the physical counterpart of CLT (which is
strictly speaking a mathematical property), as regards the statistical 
characterization of kinetic variables (velocity, momentum), the
dynamics of which is intrinsically distributional (owing to the
conservation principles) and not additive.
It can be  stated, in a pictorial way, that while ``kinematic
Gaussianity'' stems from CLT, as in the spatial propagation of Brownian motion,
``dynamic Gaussianity'', as in equilibrium velocity and momentum
distributions, is a consequence of CMT, with its imbedded supergeneric
occurrence.

The definition of  CMT finds another major application  in the study
of thermalization, and of equilibrium properties of molecular gases, in which,
apart from particle-particle collisions, quantum effects, related to the
structure of the quantized energy levels of the molecules, should be necessarility
taken into account. This problem, that is an extension of the work by
Einstein \cite{einstein1916,milonni} on the momentum transfer by emission and 
absorption
of radiation, and of the stochastic modeling of radiative effects \cite{PG_radiation} will be developed in a forthcoming work.


\begin{thebibliography}{60}
\bibitem{kaclibro} M. Kac, Statistical Independence in Probability, Analysis \& Number Theory, (Dover Publ., Mineola, 2018).

\bibitem{cltgen1} B. V. Gnedenko and  A. N. Kolmogorov, Limit Distributions for Sums of Independent
Random Variables, (Addison-Wesley, Reading, 1954).

\bibitem{cltgen2} V. V. Petrov, Sums of Independent Random Variables,
 (Springer-Verlag, New York, 1975).

\bibitem{cltgen3}  P. Billingsley, Probability and Measure,
(John Wiley \& Sons, New York, 1995).

\bibitem{cltgen4} B. V. Gnedenko and V. Yu. Korolev, {\it Random Summation}
(CRC Press, Boca Raton, 1996). 

\bibitem{stein} C. Stein,
Proc. of the Sixth Berkeley Symp.
on Math. Statist. and Probab., Vol. II: Probability theory, 1972, 583–602.
\bibitem{stein2} S. Chatterjee,  arXiv:1404.1392, 2014.
\bibitem{statphys1}  M. Kardar, {\it Statistical Physics of Particles}
(Cambridge Univ. Press, Cambridge, 2007).

\bibitem{gionaklages} M. Giona, A. Cairoli and R. Klages,
  J. Phys. A {\bf 55} (2022) 475002.

\bibitem{einstein} A. Einstein, Investigations on the Theory
of Brownian Movement, (Dover Publ., Mineola, 1956).

\bibitem{fractal1} S. Havlin and D. Ben-Avraham, Adv. Phys. {\bf 36} (1987) 
695.
\bibitem{fractal2} R. Klages, G. Radons and I. M. Sokolov (Eds.),
Anomalous transport, (Wiley-VCH Verlag, Weinheim, 2008).
\bibitem{super1} M. F. Shlesinger, B.  J. West and J. Klafter J,
 Phys. Rev. Lett. {\bf 58},  (1987) 1100.

\bibitem{super2} V. Zaburdaev, S. Denisov and J. Klafter,
  Rev. Mod. Phys. {\bf 87}, (2015) 483.

\bibitem{levy}  P. L\'evy, Calcul d\'es Probabilit\'es,
(Gautier-Villars, Paris, 1925).

\bibitem{alpha}  V. V. Uchaikin and V. M. Zolotarev,
Chance and Stability: Stable Distributions and their Applications,
(de Gruyter, Berlin, 1999).


\bibitem{relativity} W. Pauli, Theory of Relativity, (Dover Publ., Mineola, 1981).
\bibitem{boltz1} S. Chapman and T. G. Cowling, The Mathematical  Theory
of Non-Uniform Gases, (Cambridge University Press, Cambridge,  1952).
\bibitem{boltz2} S. Harris, An Introduction to the Theory of the
Boltzmann Equation, (Dover Publ., Mineola, 2004).
\bibitem{gibbs} P. Ehrenfest and  T Ehrenfest, The Conceptual Foundations of the Statistical Approach in Mechanics, (Dover Publ., Mineola, 2014).


\bibitem{replica} M. Mezard, G. Parisi and M. A. Virasoro,
Sping Glass Theory and Beyond, (World Scientific, Singapore, 1987).
\bibitem{jona}  G. Jona-Lasinio,  Il Nuovo Cimento   {\bf 26} (1975)  99.

\bibitem{superuniv} P. Goswami and S. Chakravarty, Phys. Rev. B {\bf 95} 
(2017) 075131.
\bibitem{dunkel1}  J. Dunkel and P.    H\"anggi,  Phys. Rep. {\bf 471} (2009)
 1.

\bibitem{dunkel2} J. Dunkel, P. Talkner and P. H\"anggi,
New J. Phys. {\bf 9} (2007) 144.
\bibitem{giona_rel} M. Giona, EPL {\bf 126} (2019) 50001.
\bibitem{einstein1916} A. Einstein,  Phys. Zeit. {\bf 18} (1917)  121.

\bibitem{milonni} P. W. Milonni, The Quantum Vacuum. An
introduction to Quantum Electrodynamics, (Academic Press, San Diego, 1994).
\bibitem{PG_radiation} C. Pezzotti and M. Giona, Particle-photon radiative interactions and thermalization, (2022) in preparation.
\end{thebibliography}
\end{document}